%Paper: hep-th/9409089
%From: susskind@dormouse.stanford.edu (Leonard Susskind)
%Date: Thu, 15 Sep 94 15:14:18 GMT-0800
%Date (revised): Tue, 27 Sep 94 17:12:55 GMT-0800

%%%%%%%%%%%%%%%% READ THIS %%%%%%%%%%%%%%%% READ THIS %%%%%%%%%%%%%%
%%%%%%%%%%%%%%%%%%%%%%%%%%%%%%%%%%%%%%%%%%%%%%%%%%%%%%%%%%%%%%%%%%%%
% This paper has five figures appended in a second part as a       %
% uuencoded compressed tar file with instructions for unpacking.   %
% They will be automatically included in the text if you have a    %
% functioning epsf.tex.  If you don't have that macro package      %
% (available from hep-th), or don't have the figure files,         %
% COMMENT OUT THE FOLLOWING LINE:                                  %
\input epsf
%                                                                  %
% Then, find the comment lines FIGURES HERE and comment out the    %
% lines indicated.  These lines must include the following:        %
% {\centerline{\epsfsize=3.0in \hskip 2cm \epsfbox{figX.ps}}}      %
% (There are three of them)                                        %
% The figures will probably not print correctly without epsf.tex   %
%%%%%%%%%%%%%%%%%%%%%%%%%%%%%%%%%%%%%%%%%%%%%%%%%%%%%%%%%%%%%%%%%%%%
\input phyzzx
\overfullrule=0pt
\hsize=6.5truein
\vsize=9.0truein
\voffset=-0.1truein
\hoffset=-0.1truein

\rightline{SU-ITP-94-33}
\rightline{September 1994}
\rightline{hep-th/9409089}

\vfill

%
% *** title ***
%
\title{The World as a Hologram}

\vfill

%
% *** author list ***
%
\author{Leonard Susskind}\foot{susskind@dormouse.stanford.edu}

\vfill

\address{Department of Physics \break Stanford University, Stanford,
CA
94305-4060}

\vfill

%
% *** abstract ***
%
\abstract{\singlespace According to 't Hooft  the combination of
quantum mechanics and gravity requires the three dimensional world to
be an image of data that can be stored on a two dimensional
projection much like a holographic image.

\singlespace The two dimensional description only requires one
discrete degree of
freedom per Planck area and yet it is rich enough to describe all
three
dimensional phenomena. After outlining 't Hooft's proposal I give a
preliminary informal description of how it may be implemented. One
finds a
basic requirement that particles must grow in size as their momenta
are
increased far above the Planck scale. The consequences for high
energy
particle collisions are described.

\singlespace The phenomena of particle growth with momentum was
previously
discussed in the context of string theory and was related to
information
spreading near black hole horizons. The considerations of this paper
indicate that the effect is
much more rapid at all but the earliest times. In fact the rate of
spreading is found to saturate
the bound from causality.  Finally we consider string theory as a
possible realization of 't Hooft's idea.  The light front lattice
string model of Klebanov and Susskind is reviewed and its
similarities with the holographic theory are demonstrated. The
agreement between the two requires unproven but plausible assumptions
about the nonperturbative behavior of string theory.  Very similar
ideas to those in this paper have long been  held by Charles Thorn.

}

\vfill

%
% *** PACS categories ***
%
PACS categories: 04.70.Dy, 04.60.Ds, 11.25.Mj, 97.60.Lf
\vfill\endpage

%
% *** References
%
\REF\thooftone{G 't Hooft, {\it Dimensional Reduction in Quantum
Gravity,} Utrecht Preprint THU-93/26, gr-qc/9310006 }

\REF\thorneone{C. Thorn Published in Sakharov Conf on Physics,
Moscow,
(91):447-454}

\REF\lennylorentz{L.~Susskind \journal Phys. Rev. & D49 (94) 6606.}

\REF\Bekensteinone{J. D. Bekenstein \journal Phys Rev &D49 (94) 1912}

\REF\klebanov{I. Klebanov and L. susskind \journal Nucl. Phys. & B309
(88) 175}

\REF\MTW{C Misner, K. Thorne and J. Wheeler Gravitation (1970) WH
Freeman and Co.}

\REF\lightfrontreview{J. Kogut and L. Susskind \journal Physics
Reports & 8,no 2
(73) 75}

\REF\wilsonperry{R. Perry and K. Wilson \journal Nucl. Phys. & B403
(93) 587.}

\REF\wilsonkogut{J. Kogut and K. Wilson \journal Physics Reports &
12C (1974) 75}

\REF\lipatovmueller{L. Lipatov \journal Jetp Lett & 59 (94) 596,
A.H.Mueller CU-TP-640, n.d. (94) hep-th/9408245}

\REF\feynman{R. P. Feynman, Third Topical Conference in High Energy
Collisions of Hadrons, Stoney Brook,N.Y. Sept 1969}

\REF\bjorkenone{J.D. Bjorken and E. Paschos, \journal Phys.
Rev. & 185 (69) 1975}

\REF\bjorkentwo{J. D. Bjorken, Conf, Intern. on Duality and Symmetry
in Hadron
Physics, Tel Aviv , (1971)}

% Kogut and Susskind "Scale Invariant parton Model"
\REF\scaleinvparton{J. Kogut and L. Susskind \journal Phys Rev &D9
(74) 697,3391}

\REF\alterparis{G. Altarelli and G Parisi \journal Nucl. Phys. & B126
(77) 298}

\REF\klebanovtwo{M. Karliner, I Klebanov and L. Susskind \journal
Int. Journ.
Mod. Phys.
 & A3 (88)}

\REF\larusand{A.~Mezhlumian, A.~Peet, and L.~Thorlacius, {\it String
Thermalization Near a Black Hole Horizon,} preprint SU-ITP-94-4,
NSF-ITP-94-17, February 1994, hep-th/9402125.}

\REF\veneziano{G. Veneziano, DST workshop on Particle Physics:
Superstring Theory, Kanpur 1987, ed. H.S.Mani and R. Ramachandran
(World Scientific, Singapore) p.1; Superstring '89 Workshop, Texas
A\&M University, ed. R. Arnowitt et al. (World Scientific, Singapore)
p. 86, and references therein.}

\REF\preskillthooft{Private communication, J. Preskill and G. 't
Hooft}

% Page, "Average Entropy of a Subsystem"
\REF\page{D.~N.~Page \journal Phys. Rev. Lett. & 71 (93) 1291.}

%
% *** Main body of paper ***
%

%
% *** Chapter 1 ***
%
\chapter{Counting Degrees of Freedom}

Most physicists believe that the degrees of freedom of the
world consist of fields filling space. Some theorists believe that
this may be an
overestimate and that a small distance cutoff will be required in
order to make
sense of quantum gravity. According to this philosophy the world is
about as
rich in structure as a three dimensional discrete lattice theory with
a spacing
of order the the Planck length. In this paper I will follow 't Hooft
and argue for a far more
radical decrease in the number of degrees of freedom. Instead of a
three
dimensional lattice, a full description of nature requires only a two
dimensional lattice at the spatial boundaries of the world. In a
certain sense the world is two
dimensional and not three dimensional as previously supposed. The
view I will describe combines the
ideas of Gerard 't Hooft [\thooftone ], Charles Thorn [\thorneone ]
and myself [\lennylorentz ] and
rests on the profound insights of Jacob Bekenstein concerning the
maximum entropy of a region of space
[\Bekensteinone ]. I have made an effort to keep the equations as few
and simple as possible in
order to not obscure the underlying physical concepts. The reader
will find technical
details and equations in the quoted references.

Let us temporarily suppose the world is a 3
dimensional lattice of spin like degrees of freedom. For definiteness
assume the
lattice spacing is the Planck length $l_p$ and that each site is
equipped with a
spin which can be in one of two states. For example a lattice fermion
field
theory would be of this type. Now consider the number of distinct
orthogonal
quantum states in a region of space of volume $V$.

$$
N(V) =
2^n
\eqn\stnum
$$
Where $n$ is the number of sites in $V$. The logarithm of
$N(V)$ is the maximum possible entropy in $V$ and satisfies

$$
\log N(V) = n\log 2 = {V \log 2}/{{l_p}^3}
\eqn\logn
$$

More generally one expects that if
the energy density is bounded then the maximum entropy is
proportional to the
volume of space. It is hard to avoid this conclusion in any theory in
which the
laws of nature are reasonably local. Nevertheless there is good
reason to believe that the correct
result in the quantum theory of gravity is that the maximum entropy
is proportional to  the area
and not the volume of the region. The argument is due to Bekenstein
and goes as follows.

Most of the states described in eq.\logn \ have energy so large that
a black hole would form with
a size larger than $V$. Furthermore the entropy of a black hole is
given by the Bekenstein Hawking
formula

$$
S={area\over{l_p}^2}\log 2 ={area\over{4G}}
\eqn\sbh
$$

Now suppose a region of space inside $V$ was found to have an entropy
in excess of the entropy of
a black hole just big enough to fit in $V$ but with smaller energy.
By throwing in additional
matter such black hole could be formed. Since the entropy of the
black hole would be smaller
than the original entropy the second law would be violated.
Bekenstein concludes that the maximum
entropy of the region $V$ is given not by eq.\logn\ but by eq.(1.3).
't Hooft has proposed a very
radical interpretation of Bekenstein's observation [\thooftone ].
According to 't Hooft it must be
possible to describe all phenomena within $V$ by a set of degrees of
freedom which reside on the
surface bounding $V$. The number of degrees of freedom should be no
larger than that of a two
dimensional lattice with approximately one binary degree of freedom
per Planck area. In other
words the world is in a certain sense a two dimensional lattice of
spins.

't Hooft further imagines that in the limit of a very large region
the bounding surface can be taken
to be a flat plane at infinity. In some way, the phenomena taking
place in three dimensional space
can be projected onto a distant ''viewing screen'' with no loss of
information. In what follows I
will refer to such a two dimensional surface as a ``screen'' and its
discrete lattice sites as
``pixels''. A pixel can only store one bit of information and is
therefore either lit or dark.
't Hooft has made the analogy with a hologram which stores a three
dimensional image on a two
dimensional film. As in the case of the hologram the flat two
dimensional image must be rich
enough to code the full rotationally invariant description of three
dimensional objects.

Another completely different line of reasoning leads to the
conclusion that the world is two
dimensional. Klebanov and I [\klebanov ] and independently Charles
Thorn [\thorneone ] have
discovered that string theory in the light front gauge has the form
of a two plus one dimensional
theory with no explicit mention of a longitudinal direction. In the
work of Klebanov and myself
the transverse two dimensional space is taken to be a discrete
lattice with no continuum limit
required. Thus we find similarities between string theory and 't
Hooft's holographic
idea. In this paper I will describe both viewpoints and explain their
relationship. We will find
that with one plausible assumption about the nonperturbative behavior
of super string theory,
it is an exact realization of the holographic world.

The mapping between 3+1 dimensional events and their image on the
distant screen can be
qualitatively described in semiclassical terms but the reader should
keep in mind that the full
description is necessarily quantum mechanical. We will begin with
static configurations.
I will assume that all matter is composed of elementary structureless
constituents that I will call
partons. Bits might be a better term but I will save it to denote the
unit of information. The
presence of a parton is represented on the screen by projecting it
according to the following rule.
Take light rays which pass through the screen at right angles to the
screen. Among these light rays,
one will pass through the location of the parton as in fig.[1].

\vskip 15pt
\vbox{{\centerline{\epsfysize=3.0in \hskip 2cm \epsfbox[125 370 625
700]{fig1.ps}}}
\vskip 12pt
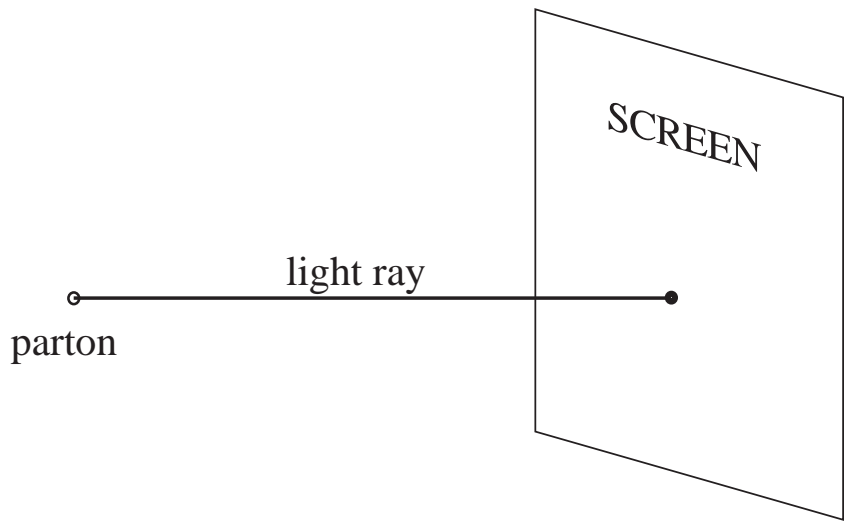
{\centerline{\tenrm FIGURE 1. A single parton being projected to the
screen}}
\vskip 15pt}

By this means the location $x_m$ is mapped to a point $X_{\perp}$. If
the screen is oriented
perpendicular to the $z$ axis then the image on the screen is a point
$[X,Y]$. We will say that
the presence of the parton is recorded by lighting the pixel at
$[X,Y]$. we want to show that no
distribution of matter will ever require more than one bit of
information per Planck area on the
screen. Let us begin with a maximally dense object, a black hole. It
will be assumed that the
entropy is found  on the horizon and that no more than one bit per
Planck area can stored there.
Now map the horizon to the screen according to the aforementioned
rule. The result is shown in
fig.[2].

\vskip 15pt
\vbox{{\centerline{\epsfysize=3.0in \hskip 2cm \epsfbox[-150 320 850
700]{fig2.ps}}}
\vskip 12pt
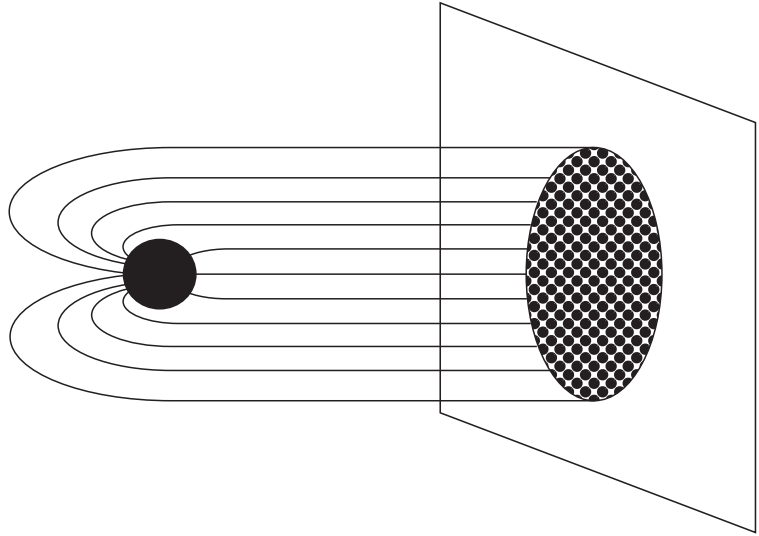
{\centerline{\tenrm FIGURE 2.A black hole projected to the screen}}
\vskip 15pt}

The image on the screen is a disc whose area is comparable to that of
the horizon. We wish to
know whether an area element on the screen is larger or smaller than
the corresponding element on
the horizon. If it is smaller, the information density on the screen
must exceed the bound of one
bit per Planck area. That this does not happen is insured by the
standard focusing theorem of
general relativity [\MTW ]. If $a$ is the area of a bundle of light
rays then the focusing theorem
states that the second derivative of ${\sqrt {a}}$ with respect to an
appropriate path parameter
is always zero or negative. Because the light rays in question all
intersect the screen parallel
to the $z$ axis the path derivative vanishes at the screen. Since the
second derivative is
negative the area element at the horizon is smaller than at the
screen.

Let us now try to increase the area density of information by hiding
a second black hole behind the
first. We will find ourselves frustrated by the lensing property of
the original black hole. We
find the image of the second hole forming a ring around the first as
in fig.[3].

\vskip 15pt
\vbox{{\centerline{\epsfysize=2.0in \hskip 2cm \epsfbox[-150 450 850
700]{fig3.ps}}}
\vskip 12pt
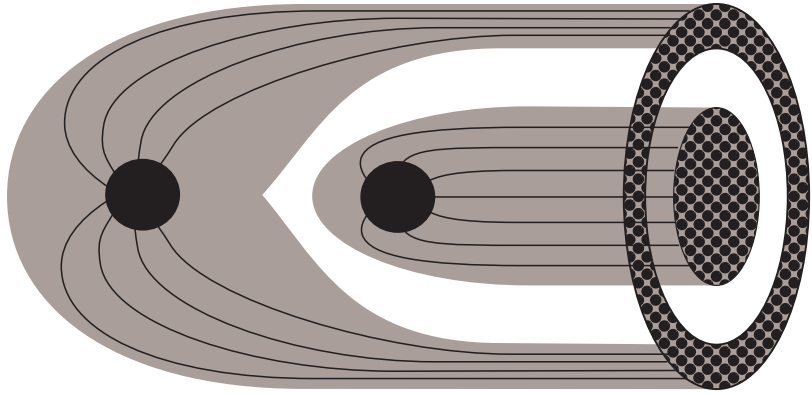
{\centerline{\tenrm FIGURE 3. You can't hide behind a black hole.
The shaded areas represent the region}
\centerline{\tenrm occupied by light rays from the screen to the
black holes.}}
\vskip 15pt}

Again the focusing theorem insures that the density in the ring does
not exceed the bound. It is
interesting to consider the sequence of images that would result from
slowly passing the first black
hole in front of the second in an attempt to eclipse it. This is
shown in fig.[4].

\vskip 15pt
\vbox{{\centerline{\epsfysize=3.0in \hskip 2cm \epsfbox[-150 320 850
800]{fig4.ps}}}
\vskip 12pt
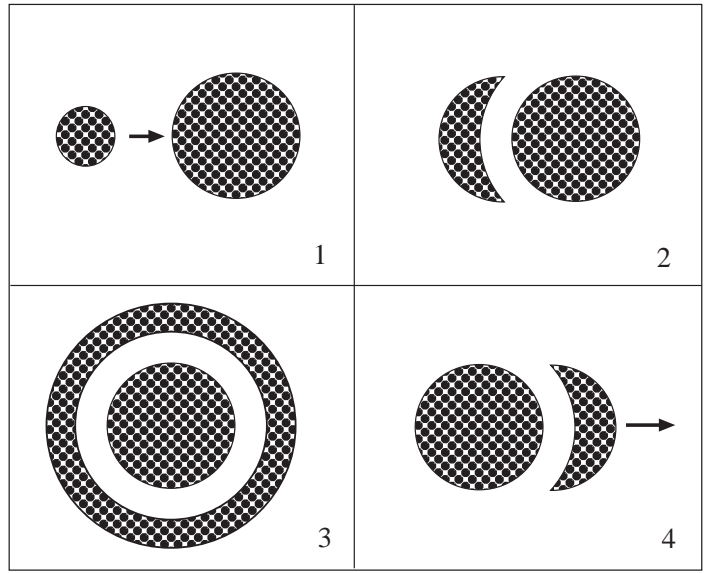
{\centerline{\tenrm FIGURE 4. One black hole passes behind another}}
\vskip 15pt}

The images behave as if they were formed of an incompressible fluid.

The image of a parton on the screen will generally not be unique in
which case
there will be more than one ray from the parton to the screen which
satisfies the criteria
for defining an image. In this case we will assume that the state of
the screen is a quantum
superposition. The image of the parton can be in any one of the
possible locations.

As another example consider a long column of length $L$ and radius
${R}$. The axis of the column is
oriented along the $z$ direction. We would like to fill the column
with as large an entropy as
possible at the minimum cost in energy so as to minimize the focusing
effects of its gravitational
field. Therefore we fill it with massless quanta of minimum energy.
The minimum energy of such a
quantum is of order $1/R$. Let the number of such quanta be of order
$N$. The maximum number of
bits of information that can be stored in this manner is also of
order $N$. Let us suppose that the
column density of information just saturates the bound of 1 bit per
Planck area.

$$
N/{R^2} \sim 1/{l_p}^2 \sim 1/G
\eqn\colmn
$$
The total energy in the column is

$$
E_{total} \approx R/{l_p}^2
\eqn\etot
$$
and the energy per unit length is
$$
E_{total}/L \approx R/{LG}
\eqn\elenth
$$

Assume the image of the column forms a disc with radius $\sim R$. We
wish to ask where the image
of a point behind the column would be located. In particular, does it
lie outside the dense image of
the column? To answer this, we consider a light ray originating at
the screen at a radius a bit larger
than $R$. The ray propagates under the gravitational influence of of
the column and is thus bent
toward the axis. A simple calculation shows that the ray will
intersect the axis approximately at
the end of the column. This is consistent with the requirement that
matter placed behind the column
will cast its image beyond the dense disc of the column.

Let us now relax the requirement that the three dimensional world is
static. Instead of coding a
point of space on the screen we want to code an event $(x,t)$. As
before we use light rays passing
through the event and intersecting the screen at right angles.
Consider a single instant of time
at the screen. All the perpendicular light rays which hit the screen
at that instant form a three
dimensional ''light front''. We may introduce a set of light front
coordinates by gauge fixing the
metric of space-time to have the form

$$
ds^2 = g_{+-} d{x^+} d{x^-} + g_{+i} d{x^+} d{x^i} + g_{ij}d{x^i}
d{x^j}
\eqn\metric
$$
where the components $x^i$ refer to transverse space and $(x^+ ,
x^-)$ are light like linear
combinations of the time and $z$ coordinates. Assume that at
${x^-}=\infty$ the metric has the
flat form with $g_{+-}=1 , g_{+i}=0 , g_{ij} = \delta_{ij}$. We may
identify $ x^+ = z+t , x^- =
t-z $. The screen will be identified as the surface $x^- = \infty$.
The trajectories
$$
x^+ = const
$$
$$
x^i = const
\eqn\nulge
$$
are easily seen to be lightlike geodesics and the surface $x^+ =
const$ is a light front. We will
follow the standard practice of using $x^+$ as a time coordinate when
doing light front
quantization . As we have seen it is also the time at the screen.

The mapping from 3+1 dimensional space to the screen is now
especially simple. A point $x^{\mu} =
(x^+,x^-,x^i)$ is mapped to the point $( x^+,x^i)$. Thus we arrive at
the following statement of
't Hooft's holographic principle:

In the light front quantization of quantum gravity, the theory can be
formulated with no direct
reference to the longitudinal direction $x^-$ and the transverse
space may be taken to be a
discrete lattice. The lattice is composed of binary pixels with
spacing of order the Planck
length. We may find it convenient to consider larger lattice spacing
and have a larger number of
configurations at each site but the net result should be one bit per
Planck cell.

The obvious question now arises as to how we code the longitudinal
location of a parton. Before
addressing this question it is necessary to review light front
quantization and the parton concept.
These concepts are familiar to people currently working on QCD and
originated in work on deep
inelastic scattering in the late 60's and early 70's. They are
probably not very familiar to
people working on black holes and gravitational physics
[\lightfrontreview ].

\chapter{Light Front Quantization and Partons}

In this section I will review some very basic concepts of light front
quantization for
conventional field theories such as QCD. For simplicity I will write
the formulae for scalar
$\phi^3$ theory. Light front coordinates are introduced as in eq
\metric\ but with the metric
chosen flat. The coordinate $x^+$ is used as a time coordinate. The
light like surfaces $x^+ =
const$ play the role of instantaneous spatial hypersurfaces. Partons
such as quarks or gluons are
described by a transverse position $X^\perp$ or transverse momentum
$Q_\perp$ and a
positive longitudinal momentum $p_-$. In addition a parton may carry
internal quantum numbers and a
spin degree of freedom which I will suppress.

Light front creation and annihilation operators $a^+(X^\perp,k_-)\;\;
,\;\; a^-(X^\perp,k_-)$
create and annihilate particles in the usual way. The naive
(classical) free Hamiltonian has the
form

$$
H_0\;=\;\int {d^2 P_\perp}{dp_-}\;{{a^+(p){[{P^2_\perp}+M^2}]a^-(p)}
\over 2p_-}
\eqn\hamo
$$
where M is the mass of the parton. Classical interactions take the
form

$$
H_I\;=\;\lambda \int
{d^3p\;d^3q}{{{a^+(p)a^+(q)a^-(p+q)}F(p_-/q_-)}\over
\sqrt[({p_-})({q_-}){(p_-+q_-)}]}\;+h.c.
\eqn\hami
$$
where the functions $F$ are simple rational functions that depend on
the spins of the particles
involved. For scalar particles they are constants. In addition there
may also be terms with higher
powers of the creation and annihilation operators.

The operators
$a^{\pm}$ have canonical commutation or anti commutation relations

$$
[a^-(p)\;,\;a^+(q)]\;=
\;\delta^2({P_\perp}-{Q_\perp})\;\delta(p_--q_-)
\eqn\cancom
$$

 We will be especially interested in the properties of the theory
under longitudinal boosts which
act according to
$$
\;\;p_-\to e^{\omega}p_-
$$
$$
P_\perp \to P_\perp
\eqn\ptran
$$
where $\omega$ is the hyperbolic boost angle.

The operators $a^{+}\;,\;a^-$ naively transform as
$$
a(P_{\perp},p_-) \to e^{\omega \over 2}a(P_{\perp},{e^\omega}p_-)
\eqn\atran
$$

The Hamiltonian transforms as
$$
H\;\to\;{H}e^{-\omega}
\eqn\htran
$$
An important advantage of the light front method is that the vacuum
is the naive Fock space
vacuum annihilated by all the $a^-$. On top of this vacuum we erect a
space of parton states by
acting with the $a^+$ operators

$$
|k_1\;k_2.....\rangle \;=\; a^+(k_1) a^+(k_2)...|0\rangle
\eqn\fok.
$$
The Hamiltonian acts in this space.

According to the most naive view of the theory the Hamiltonian is the
classical one and the
eigenvectors are well defined convergent superpositions of Fock space
states. Boosting a system
along the $z$ axis is supposed to be straightforward. Each parton
momentum transforms according to
eq.\ptran. The transformation simply acts as a scale transformation
of the longitudinal momentum
axis. According to this naive view a longitudinal boost preserves the
transverse dimensions of a
system and rescales or Lorentz contracts all longitudinal dimensions.
However, the correct
description is far more complicated due to various kinds of
divergences which occur in the theory
[\wilsonperry ].

The boost problem has a close analogy with the problem of scale
invariance in ordinary quantum field theory. Classical scale
invariance will usually be destroyed by divergent high frequency
effects and even if it is possible to preserve it , the invariance
is realized in some anomalous form. For example, fields will have
anomalous dimensions which will cause them to transform in a
noncanonical manner. Before discussing boosts I will review the
subtleties of
scale invariance in the language of Hamiltonian quantum
mechanics.

The degrees of freedom are the spatial Fourier modes $\phi(p)$
where the spatial momentum $p$ is any real 3 vector. For our purposes
it will be important to  cut the
theory off in the infrared. This may be done by simply eliminating
modes with momenta less than
some cutoff $\kappa$. Alternatively we could put the system in a
finite box of size $\kappa^{-1}$.

Suppose the classical theory has scale invariance under which all
momenta including the infrared regulator $\kappa$ rescale

$$
p \;\to\;{\lambda}{p }
\eqn\scale
$$
The naive scale invariance would imply the following:

1) The spectrum of the Hamiltonian rescales under \scale. Each
eigenvalue of $E_i$ maps to a rescaled value.

$$
{E_i}\;\to\; {\lambda}{E_i}
\eqn\rescale
$$

2) The wave functionals of the eigenvectors transform in a naive
way.
$$
{\Psi_i[\phi(p)]} \;\to\; {\Psi_i[\lambda\phi(\lambda {p})]}
\eqn\wavescale
$$
Each fluctuation of wave number $p$ simply stretches to wave number
$\lambda {p}$.

The problem with the naive view is that the low and high
frequencies are not at all decoupled [\wilsonkogut ]. This causes
divergences in
integrations over large momenta which make everything infinite. To
define the theory an ultraviolet cutoff must be supplied. From an
operational point of view the use of an ultraviolet cutoff is
always correlated with a particular experimental setup. A given
apparatus will be sensitive to frequencies up to some maximum $\nu$.
Beyond that frequency the apparatus cannot see. For example, an
ordinary particle
detector cannot detect the
very frequent and rapid fluctuations of baryon number in the vacuum
that take place with GUT scale frequency. On the other hand if the
detector has internal processes of sufficiently high frequency, then
these fluctuations become
visible. For example the detector could contain a vessel with
radiation at GUT scale temperatures.
This connection between apparatus and ultraviolet frequency cutoff is
fundamental and should be kept
in mind in what follows.

The mathematical description of the cutoff theory will not have
modes with momentum higher than the ultraviolet cutoff $\nu$. It
will have a Hamiltonian $H_{\nu}$, energy levels $E_{\nu}$, and wave
functionals $\Psi_{\nu}$. Now consider a scale transformation by
which I mean a rescaling of
the infrared cutoff, leaving the ultraviolet cutoff fixed. In
general the energy levels will not simply rescale and we will say
that scale invariance is broken. There may however be special
Hamiltonians that I will call ''fixed point Hamiltonians'' that
preserve certain features of the scale invariance. In particular for
such Hamiltonians, whenthe infrared cutoff $\kappa$ is rescaled the
energy spectrum
scales just as in \rescale. Such a theory is said to be scale
invariant.

Although the energies transform naively in fixed point theories the
wave functionals do not. The transformation properties are far more
complex than \wavescale. A fluctuation or bare quantum may
transform into a superposition of several bare quanta. To see why
this is so it is helpful to think of the rescaling operation in
another equivalent way. Instead of decreasing the infrared cutoff
$\kappa$ while holding fixed the ultraviolet cutoff $\nu$, we
increase $\nu$, while holding $\kappa$ fixed. In a fixed point
theory these are equivalent because things only depend on the ratio
of the two cutoffs. However the perspective is a bit different.
When we increase $\nu$ we uncover new degrees of freedom at short
distances which were previously not part of the description. Bare
field quanta of the more coarse grained description are found to
have structure at the new scale. A bare gluon is found to have an
amplitude to be a closely spaced quark pair while a quark has an
amplitude to contain a nearby gluon. For this reason the action of
the dilatation generator on the Schrodinger picture wave functional
is generally very complicated and theorists
avoid talking about it. However the problems resurface when trying
to unravel the intricacies of very high energy scattering . Now let
us
return to the problem of boosts.

The momentum space integrals of light front perturbation theory can
diverge at high frequency in
two ways. The most obvious way is at large values of the transverse
momentum.  These divergences
reflect the usual short distance singularities that are familiar from
covariant perturbation
theory. We will return to these divergences but for the moment let us
temporarily suppress them by
assuming that transverse momentum integrals are finite and dominated
by values of $P_{\perp}$
of order some characteristic scale in the problem.  For example in
QCD the confinement scale of 300
mev is natural. In the quantum theory of gravity the Planck mass
would be the appropriate scale. The
characteristic momentum scale will be denoted $M$.

The second form of high frequency divergence that can occur involves
integrals over $p_-$. The
range of such integrals is always restricted to values of $p_-$ which
are less than the total
longitudinal momentum $p_-(tot)$. From the form of eq.\hamo\ we see
that the energy of a parton
diverges when $p_-$ tends to zero. This region describes large
distances in the $x_-$ coordinate
but small distances in $x^+$. The divergences in this region are not
connected with scale
transformation anomalies but rather anomalies in the behavior of
Lorentz boosts.

An example of such a divergence involves the probability for a single
physical particle of momentum $({p_-}+{q_-})$ to be a pair  of
partons of momentum $p_-$ and $q_-$. In second order perturbation
theory it is given by

$$
\int {dq_-}|{\langle p+q|H_I|p,q \rangle}|^2 [E(p+q)-E(p)-E(q)]^{-2}
\eqn\prob
$$
Using \hami we find that the integrand behaves like

$$
{q_-}F^2(p_-/q_-)
$$
as $q_-\;\to\;0$.
Furthermore, if the low momentum parton has spin J, the function $F$
has the form $(q_-)^{-J}$ in this limit. Evidently the probability
diverges for $J \ge 1$. These divergences indicate that the
population of partons
can become infinite at low longitudinal momentum thus invalidating
the naive picture.

It is useful to imagine the degrees of freedom $a^{\pm}(p)$ being
laid out on the $p_-$ axis. The
length of the axis is finite and given by $p_-(tot)$. A longitudinal
boost is represented as a
scale transformation of this axis. Lorentz invariance requires
perfect scale invariance with the
Hamiltonian transforming according to eq.\htran. The divergences at
low $p_-$ can potentially
disrupt the classical invariance in a manner similar to the way
ordinary divergences can ruin
scale invariance. As in that case  we must introduce a cutoff. This
time the cutoff is on low
values of $p_-$. Thus we introduce a cutoff which eliminates all
$p_-$ less than
$\epsilon$  and search for a fixed point of the renormalization
group. Such fixed points
classify the possible boost invariant theories. The problem of
finding and understanding
such fixed points in gauge theories is extremely complicated and is
equivalent to understanding
the behavior of matter under extreme boosts [\lipatovmueller]. This
is not the place for a
technical review of the work currently taking place in this area  but
we will need to understand
some of the basic logic.

 We begin by imagining that we have a set of detectors or apparatuses
that are sensitive to
frequencies up to some maximum of order $M^2{\epsilon}^{-1}$. An
appropriate description  should be possible with longitudinal momenta
cutoff at
$p_-=\epsilon$. It is important to keep in mind this connection
between the cutoff procedure and an
apparatus.

The system with cutoff $\epsilon$ is described by a Hamiltonian
$H(\epsilon)$ which will not in
general have the classical form. Let us fix the total longitudinal
momentum to be $p_-(tot)$. The
Hamiltonian then has a set of energy levels $E_i(\epsilon)$ and wave
functions $\Psi_i(\epsilon)$.
Longitudinal boost invariance or fixed point behavior requires the
energies to be independent of $\epsilon$ but says nothing simple
about the wave functions.

Now consider boosting the total momentum  (equivalently decreasing
$\epsilon$) by a factor
$\lambda$. Each energy level must map into a new level with energy
${E_i}\over \lambda$. Again, in
general nothing simple can be said about the wave functions except
that they should be obtained by
some unitary mapping from the original wave functions. This mapping
is the boost operator. It
should be borne in mind throughout that the boost operation really
represents the relative boost
between a system and an apparatus with a resolution time of order
$\epsilon$.

The problem of determining fixed points and the behavior of the wave
functions under boosts in
QCD is not a well developed subject [\wilsonperry ]. I will describe
some simple possible behaviors of
the boost operation that have been discussed in the past. These
behaviors do not represent
established fixed points but are simplified models. The real behavior
of a theory like QCD is
probably a good deal more complicated.

 1) The Einstein Lorentz Fixed Point

   In very simple field theories with no divergences the wave
function of a physical particle or
system of particles is a convergent Fock space state with a finite
average number of partons. If
the cutoff $\epsilon$ is sufficiently small the probability to find a
parton with $p_-<\epsilon$
is vanishingly small. Boosting the system is trivial. Each parton
shifts to its boosted position.
Transverse sizes are invariant and longitudinal sizes contract. Most
likely no real 3+1 dimensional
quantum field theory works this way.

2) The Feynman Bjorken Fixed Point

In the case of gauge theories longitudinal divergences induce a
divergent distribution of
partons at low $p_-$. According to the Feynman Bjorken model
[\feynman ], [\bjorkenone ]
[\lightfrontreview ], the number of partons per unit
$p_-$ behaves like
$$
{dn\over{dp_-}} \sim {1\over{p_-}}
\eqn\partdis
$$
Introducing a high frequency $(p_-<\epsilon)$ cutoff is more serious
this time since it eliminates
degrees of freedom which are present in the hadron. In this case when
a boost doubles the
longitudinal momentum of each parton, a hole is left in the region
$\epsilon<p_-<2\epsilon$. New
partons must be added to fill this region. In other words the boost
operator must contain an extra
term which acts as a source of partons of low $p_-$. These are the
partons that Feynman
called the ''wee'' partons. The wee partons create interesting
anomalies in the behavior of
matter under boosts. For example, because they always carry low
longitudinal momentum they
contribute a cloud which does not Lorentz contract [\bjorkentwo ].
Furthermore they tend to be
found at progressively larger transverse distances from the center of
mass. For example in simple
Regge approximations their average distance satisfies
[\lightfrontreview ]
$$
R^2_{\perp}(wee) \sim log{{p_-(tot)}\over {\epsilon}}
\eqn\herewee
$$

In the Feynman Bjorken model the quantum numbers of a hadron are
carried by ''valence'' partons
which carry finite fractions of the total momentum and therefore
behave as in the Einstein Lorentz
case. Thus the spatial distributions of charge, baryon number and
angular momentum Lorentz contract and
do not transversely spread. This is necessary in a conventional field
theory in order that the local
current operators be well defined.

The cutoff $\epsilon$ is arbitrary but just as in ordinary
renormalization theory, physical
applications may make a particular choice most convenient. In general
a given experiment will be
sensitive to frequencies up to some maximum. For example in the case
of a high energy particle colliding with
a fixed target the target determines some range of frequencies that
it is sensitive to. Retaining higher frequencies in the description
only
leads to unnecessary complexity. The
consequences of eq.\herewee\  in such a case would be logarithmic
increase of the cross section
with energy. We see here the interplay between the behavior of the
boost operation, the low $p_-$
partons and very high energy scattering amplitudes.

3) The K.S. Fixed Point

   Divergences at large transverse momentum can alter the boost
operation. To understand this we
must generalize the cutoff procedure so that we cutoff all modes with
frequency greater than some
value $\nu$
$$
{{{K^2_\perp}+m^2}\over{2k_-}}>\nu
\eqn\cutoff
$$
Once again we can eliminate the effects of the cutoff by either
letting ${\nu \to 0}$ or $p_-\to
\infty$. Let us keep $\nu$ fixed. The phase space for partons is
shown in fig[5].

\vskip 15pt
\vbox{{\centerline{\epsfysize=3.0in \hskip 2cm \epsfbox[-150 320 850
700]{fig5.ps}}}
\vskip 12pt
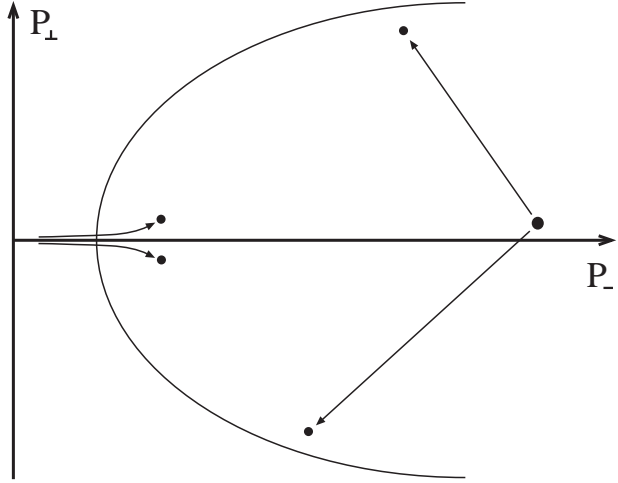
{\centerline{\tenrm FIGURE 5. Parton phase space}}
\vskip 15pt}

Consider a low longitudinal momentum parton. The interaction terms in
$H$ allow that parton to
split into two partons, each with smaller $p_-$ than the original
[\scaleinvparton ]
[\alterparis ]. From fig [5] it is obvious that the transverse
momentum of the pair must be small.
This insures that the probability for the original parton to to
become a pair is small (now
ignoring divergences at low
$p_-$). Now boost the system so that the parton has a much larger
$p_-$. The transverse phase
space for it to split is much bigger. If the theory has transverse
divergences then that
probability will become large as the system is boosted. Eventually
the parton will be replaced by
two or more partons closely spaced in transverse space. The effect
continues as the system is
boosted so that the partons reveal transverse fine structure within
structure ad infinitum. Thus
the K.S. fixed point boost operator contains terms which continuously
create wee partons which
migrate to larger $p_-$. As they do so they split into short distance
pairs which continue to
migrate and split. The new consequence for high energy scattering is
the existence of processes
involving large transverse momentum jets.

4) The Free String Fixed Point

I will now describe a behavior that is not found in ordinary field
theory but occurs in string
theory. Before discussing real string theory it is interesting to
describe it in parton terms
[\lennylorentz ]. Let us start with a single parton of longitudinal
momentum of order $\epsilon$.
Now boost it to twice its original momentum. Instead of finding a
parton of twice the momentum we
find two partons, each of the original momentum $\epsilon$. The two
partons have a transverse wave
function that is rotationally symmetric and is peaked at a transverse
radial distance $l_s
=\sqrt{\alpha^\prime}$ where $\alpha^\prime$ is the usual dimensional
string constant. Now double the
momentum again. Each of the two partons is resolved into two more
with similar wave function.
Unlike the K.S. case the evolution does not create pairs of smaller
and smaller transverse size.
Moreover no parton is ever found with $p_->\epsilon$. All partons are
wee. The idea that string
theory is a parton theory of wee partons has also been emphasized in
the past by Charles
Thorn[\thorneone ].

After $n$ iterations the total number of partons is $2^n$ and the
total longitudinal momentum is

$$
p_-(tot) ={\epsilon}2^n
\eqn\peetot
$$

It is also easy to show that after many iterations the transverse
density of partons relative to
the center of mass is gaussian with radius satisfying

$$
R^2_\perp ={l^2_s}n={l^2_s}log({p_-}/\epsilon)
\eqn\slogro
$$
We might also expect no Lorentz contraction to take place since all
partons are wee.

The implications of such a parton model for high energy cross
sections are interesting.  Let a
particle collide with a fixed target. Also suppose that the
scattering is weak enough that the
total cross section is additive in the constituent partons. Since the
number of partons is
proportional to the momentum of the incident particle the cross
section grows linearly with
laboratory energy. However, this behavior together with eq.\slogro\
can not persist indefinitely
without violating unitarity. Either the addition of partons will
eventually lead to shadowing
corrections or the geometric size of the parton cloud will have to
grow more quickly. As we shall see
the latter option is correct.

Now let us compare the string-like parton model with real string
theory in the light front frame.
I will work in units in which $l_s=1$. A free string is described by
a transverse coordinate
$X_\perp(\sigma,x^+)$ which is a function of the parameter $\sigma$
and the light front time $x^+$.
It is convenient to allow
$\sigma$ to run from zero to
$p_-(tot)$. A longitudinal boost is then seen to be a rescaling of
$\sigma$. The operator $X_\perp$
can be expanded in normal modes $X_l$ and the center of mass
$X_\perp(cm)$

$$
X_\perp(\sigma , x_-) -X_\perp(cm)= \sum_1^{\infty} \left \{ X_l
\exp{{il(\sigma - x^+)}\over p_-} + {\tilde X}_l \exp{{il(\sigma +
x^+)} \over {p_-}} \right \}
\eqn\norm
$$

Consider the mean square size of the string
$$
R^2_\perp=\langle[(X_\perp(\sigma)-X_\perp(cm)]^2\rangle
\eqn\size
$$
The matrix element gets a contribution from each mode and the result
diverges

$$
R^2_\perp=\sum_1^\infty{1\over l}=log(\infty)
\eqn\toobig
$$
The problem is that we have not introduced a high frequency cutoff.
{}From \norm we see that the
frequency of the $lth$ mode is $l/p_-$. The highest allowable
frequency when the cutoff is in
place is of order $1\over \epsilon$ so that the highest allowable
mode is
$$
l_{max} = {{p_-}\over \epsilon}
\eqn\lmax
$$
The logarithmic divergence is now replaced by \slogro

To compute the longitudinal size of a string is more subtle . String
theory does
not endow the string with an independent longitudinal coordinate. The
coordinate $x^-(\sigma)$ is
determined in terms of the transverse coordinates. The longitudinal
size was computed in
[\lennylorentz ] where it was found to also diverge if no high
frequency cutoff was used. This
time the divergence is quadratic and gives
$$
R^2_- ={1\over {\epsilon^2}}
\eqn\rlong
$$
The longitudinal size does not Lorentz contract as $p_-$ increases.

One other feature of the string wave function involves the length of
the transverse projection of the string as $p_-$
increases. It was shown in ref[\klebanovtwo ] that as the number of
modes of the string increases the
transverse length of string increases linearly. This means that the
string length is
proportional to $p_-$. To understand the connection with the free
string parton model we picture
the string as made up of string bits of length
$l_s$. We see that the number of bits or partons is proportional to
$p_-$ and each bit is wee.

The scale invariance associated with boosts is familiar in string
theory in another form. According to the usual rules of light front
string theory the total range of the parameter $\sigma$ is from zero
to $p_-(tot)$. Therefore a boost is realized as a rescaling of
$\sigma$. In other words the boost-scale invariance is equivalent to
invariance with respect to world sheet scale transformations. As is
well known, this invariance places strong restrictions on the
possible backgrounds in which strings can propagate. Among these
restrictions are Einstein's, and Maxwell's equations as well as Yang
Mills equations..

Before returning to the discussion of the holographic world I want to
describe one more idea that
originated in attempts to do numerical QCD work using light front
methods. It involves a method
of regulating the low $p_-$ divergences. It consists of replacing the
infinite $x_-$ axis by a
periodic box of size $1/\epsilon$. This has the effect of replacing
the $p_-$ axis by a
discrete lattice with spacing $\epsilon$. The smallest allowable
longitudinal momentum is
$\epsilon$. This method of regularization will prove particularly
convenient in the next section.

\chapter{The growth of Particles With Momentum}

In order to code the longitudinal motion	of systems on the
screen we
will assume that
longitudinal momentum comes in discrete units of size $\epsilon$.
Eventually, to get
exact results $\epsilon$ should tend to zero but as for ordinary
cutoffs it is usually
more convenient to choose it according to some physical criterion
involving an experimental situation.

Consider a single pixel at transverse position $X_\perp$. Since a
pixel can record only a single
bit of information it can at most record the presence or absence of a
parton but not its state of
longitudinal motion. Therefore we will assume a lit pixel at
$X_\perp$ represents a parton of
minimal longitudinal momentum $\epsilon$. Now increase the momentum
to twice the minimum. Since a
pixel can not be lit twice we are forced to light two pixels. In
general a system with momentum
$p_-=N{\epsilon}$ will be identified with N lit pixels. A system is
boosted to large momentum not
by boosting its partons but by increasing their number. In other
words all partons are wee and no
two of them may occupy the same pixel.

The relationship between this behavior and the free string fixed
point described in the last
section is straightforward. Crudely speaking when the momentum is
doubled each parton must be replaced
by two with a average separation which we call $l_s$. That scale
could be as small as the Planck scale
$l_p$ or it could be larger. The ratio is a dimensionless number $g$
$$
l_p=gl_s
\eqn\ratio
$$
If $g$ is very small the evolution of the parton distribution as it
is boosted is, at first, unconstrained
by the condition that the density not exceed $1\over {l^2_p}$ and it
develops according to
the free string behavior. However, since the number of partons
increases like $p_-$ but the radius
of the distribution increases only logarithmically, a point will come
at which the density becomes
$l^{-2}_p$. At this point the area of the system must begin to
increase more rapidly. Eventually
the area occupied by the particle must grow like
$$
A\sim {l_p}^2{p_-\over \epsilon}
\eqn\fastgro
$$

The constant $g$ is nothing but the string interaction coupling
constant. No matter how small it
is, interactions will become important when the number of modes
becomes so large that the density
gets to be $g^{-2}$ in string units. The basic requirement for
consistency with the bound on
area density is that the effect of interaction is so repulsive that
the partons create an
incompressible fluid.

The prediction that the size of a transplankian particle should grow
so rapidly with momentum at first
seems nonsensical, especially since the coefficient involves the
parameter $\epsilon$. To see that it
is not absurd we consider a gedanken experiment to measure the radius
of a high energy particle
(projectile). The projectile is allowed to collide with a fixed
particle (target) at an impact
parameter $b$. The momentum of the projectile will be varied but the
target is to be kept fixed.
Any momentum dependence in the behavior of the collision will be
attributable to variations in
the properties of the particle.

Strictly speaking the collision cross section is infinite because of
the long range tail of the
gravitational field. However if the impact parameter is very large
the collision will result in a
very weak nudge of the target. We will ignore such gentle barely
detectable events. If the impact
parameter is very small, a highly  inelastic collision may result.
The maximum impact parameter
for which such a ``real'' collision will take place with significant
probability is a measure of
the size of the projectile [\veneziano ].

Although no data exists or is ever likely to exist for the
superplanckian collisions we are
considering, the answer is easily guessed from semiclassical
considerations. Let the center of
mass energy of the collision be

$$
E_{cm}=\sqrt s
\eqn\ecm
$$
In the center of mass frame both particles are relativistic.
Classical general relativity tells us
that if the impact parameter is of order

$$
b\; \sim \; G\sqrt s
\eqn\critb
$$
or smaller, a black hole will form. The formation of a black hole is
a highly inelastic event
which definitely disrupts the target.

Returning to the Laboratory frame we find that the condition for
black hole formation is

$$
b\;<\;G\sqrt({E_T}\;{p_{lab}})
\eqn\blab
$$
where $E_T$ is the energy of the target and $p_{lab}$ is the
projectile momentum in the lab frame.
Evidently the area of the projectile grows at least as fast as
$$
A\;\approx {{l_p}^4}{E_T}{p_{lab}}
\eqn\area
$$
Thus it is seen that the prediction that the area grows as $p_-$ is
borne out. Furthermore by
comparing \area with \fastgro we find that the role of the parameter
$\epsilon$ is played by
$$
\epsilon\;=\;{1\over{{E_T}{l_p}^2}}
$$
In fact it is not surprising that the ability of a target apparatus
to detect high frequencies
should be limited, through the uncertainty principle, by its energy.
The same correspondence
between target energy and resolution time is also found in weakly
coupled string theory. In that
case an incoming projectile has an area which grows according to
\slogro. Interpreting $\epsilon$
as being inversely proportional to target energy would imply that the
interaction region grows
logarithmically with center of mass energy. This is the usual
consequence of Regge behavior.

\chapter{Boosts, Black Holes and Wee Partons}

Consider the region of space-time near the horizon of a massive black
hole. To a very good
approximation it may be replaced by flat Minkowski space-time. In
terms of ordinary cartesian
coordinates the metric is given by

$$
ds^2\;=\;dT^2-dZ^2-dX^idX^i
\eqn\mink
$$
While in terms of schwarzschild coordinates the same space-time is
given by

$$
ds^2\;=\;{({{dt}\over {4MG}})^2}{\rho}^2-d{\rho}^2-dX^idX^i
\eqn\rind
$$
In \rind we have used the coordinate $\rho$ which denotes proper
distance from the horizon instead
of the usual radial coordinate $r$. The Minkowski and Schwarzschild
coordinates are related by

$$
Z\;=\;\rho {\cosh(t/4MG)}
$$
$$
T\;=\;\rho {\sinh(t/4MG)}
\eqn\minshld
$$
The horizon of the black hole is the surface $t=\infty$.

The effect of a Schwarzschild time translation on the Minkowski
coordinates is a hyperbolic
rotation by angle

$$
\Delta \omega \;=\; {{\Delta t}\over{4MG}}
\eqn\hyprot
$$
In other words it is a boost. Following the progress of a system as
it approaches the horizon of a
black hole is tantamount to following its behavior under Lorentz
boost as the magnitude of the
boost parameter increases indefinitely. Any attempt to describe
matter from the standpoint of an
observer outside a black hole must come to grips with the issues of
the previous sections.

The same point can be made in another way. As a particle falls toward
the horizon it is
accelerated from the point of view of a static observer. The momentum
of the particle increases
like

$$
p\;\to\;exp(t/4MG)
\eqn\pgro
$$
To give a proper account of the particle from the Schwarzschild frame
we must have a description
which is valid at ever increasing momentum.

Furthermore the infalling matter is ''detected'' by target particles,
namely the outgoing thermal
Hawking radiation. In the Schwarzschild frame, near the horizon,
these particles have a fixed
energy distribution centered near the Planck energy. The situation is
essentially the same as when
a high energy particle interacts with a fixed target.

Let us therefore consider the implications of the various behaviors
described in sections 2 and 3.

1) Einstein Lorentz

The particle is described as a fixed population of partons with no
wee  partons. As the particle
falls, the Schwarzschild description sees a particle of fixed
transverse size. Longitudinally the
particle contracts exponentially rapidly in accord with \pgro.
Eventually it occupies an
vanishingly thin layer just above the horizon. This is the picture
most theorists have in mind.

2) Feynman Bjorken

Next add wee partons to the picture. As the particle approaches the
horizon we find an expanding
halo that spreads out over the horizon and refuses to Lorentz
contract. However the halo carries a
negligible fraction of the momentum. The charge, angular momentum,
isospin, baryon number, etc.
are carried as in the more naive Einstein Lorentz case.

3) K.S.

Adding the effects of large $P_\perp$ degrees of freedom we find that
the formerly point like
partons dissociate into small groups of partons, each of which
further dissociates into yet
smaller groups ad infinitum ( or until the planck or string scale is
reached.)

4) Free String

In this case as a particle falls toward the horizon its partons
dissociate into pairs. The
evolution is similar to the replication of a dilute collection of
microorganisms on a petri dish
[\larusand ]. As long as the density remains much lower than the
Planck density the area of the
distribution grows logarithmically.

$$
R^2_\perp\;=\;{l^2_s}\;\log{p_-}
\eqn\stgro
$$

Since the particle is composed entirely of wee partons it does not
contract longitudinally and
therefore it fills a nondecreasing layer above the horizon.
Furthermore the various attributes
which distinguish the particle type are spread according to \stgro.
Using \pgro  we find that the
particle grows with Schwarzschild time according to

$$
R^2_\perp\;=\;{l^2_s}{t\over {4MG}}
\eqn\slogru
$$
As explained in [\lennylorentz ], \slogru represents a sufficiently
rapid growth rate for the
particle to cover the horizon in a time short by comparison with the
evaporation time

$$
t_{spread}\;=\;{{M^3G^3}\over {l^2_s}}
\eqn\sprdslo
$$

As the particle or string grows the density at the center also grows
like the momentum. As we have
seen the density will approach the Planck density after which the
system must spread more rapidly.
Eventually the area of the infalling particle will grow in proportion
to its momentum.

$$
A\;\sim\;G{\exp(t/4MG)}
\eqn\speedy
$$
With this more rapid rate of growth the horizon is covered in a much
shorter length of time than
given in \sprdslo . The corrected formula is

$$
t_{spread}={4MG{\log(M^2G^2)}}
\eqn\sprdqwk
$$

The spreading rate predicted by \speedy appears to be closely
connected to another basic physical
principle. It is exactly the fastest rate consistent with causality
[\preskillthooft ]. To see
why we need to introduce the idea of the stretched horizon. This is a
surface a fixed proper
distance,
$\rho_0$,  from the horizon where infalling matter collects. The
numerical value of the distance
does not matter. In terms of Minkowski coordinates the stretched
horizon is given by

$$
{Z^2}-{T^2}\;=\;{\rho}^2_0
\eqn\sthz
$$
Now suppose an event occurs at some point of the stretched horizon.
With no loss of generality we
may take that point to be

$$
X^i\;=\;0
$$
$$
T\;=\;0
$$
$$
Z\;=\;\rho_0
\eqn\point
$$
The signal from this event is bounded by the forward light cone
defined by

$$
T^2\;=\;(Z-\rho_0)^2+{R^2}_{\perp}
\eqn\lc
$$
where $R^2_\perp\;=\; X^iX^i$. The intersection of the light cone
with the stretched horizon is

$$
R^2_{\perp}\;=\; {2Z\rho_0}-2\rho^2_0
\eqn\insect
$$
For large Schwarzschild time this becomes

$$
R^2_\perp\;=\;{\rho^2_0}exp(t/4MG)
\eqn\exgro
$$
which agrees with \speedy. The same exponential growth governs the
spreading of a classical
perturbation on the stretched horizon.

Thus far, I have mainly emphasized the anomalous transverse behavior
of matter under extreme boosts and the implications for black hole
physics. Let me now turn to the importance of the breakdown of
Lorentz contraction. Once again start with the Einstein Lorentz
behavior. A chunk of matter will occupy a longitudinal extension of
order its inverse momentum. Using \pgro we find

$$
{\Delta}{\rho}\;\sim\;\exp{[-t/4MG]}
\eqn\contract
$$
As seen by the observer outside the horizon, the matter contracts
into an indefinitely thin layer. For this reason, the horizon can
accommodate arbitrary amounts of Einstein Lorentz matter.

Wee partons, by contrast, are not subject to Lorentz contraction.
This implies that in the Feynman Bjorken model, the halo of wee
partons eternally ''floats'' above the horizon at a distance of
order ${10}^{-13}cm$ as it transversley spreads. The remaining
valence
partons carry the various currents which contract onto the horizon as
in the Einstein Lorentz case.

By contrast, both the holographic theory and string theory require
all partons to be wee. No Lorentz contraction takes place and the
entire structure of the string floats on the stretched horizon. I
have explained in previous articles how this behavior prevents the
accumulation of arbitrarily large quantities of information near the
horizon
of a black hole. Thus we are led full circle back to Bekenstein's
principle that black holes bound the entropy of a region of space to
be proportional to its area.

\chapter{ Lattice String theory }

Ordinary bosonic string theory has been written in a form which may
help establish the
relationship between string theory and the holographic principles
described in the previous
sections [\klebanov ][\thorneone ]. I will not reconstruct the theory
here but instead the salient
features will be listed and described. Details can be found in the
original reference.

1) The theory is a light front quantum theory. The transverse plane
is replaced by a discrete
lattice with spacing $l_s$. The lattice spacing is kept FIXED
throughout. A lower longitudinal
momentum cutoff $\epsilon$ is introduced so that $p_-$ comes in
discrete units.

2) The theory is a gauge theory. On each link of the transverse
lattice $N\times N$ matrix fields
exist. These fields have no $x^-$ dependence. The matrix fields
create and annihilate string bits
which carry $N\times N$ matrix indices.  In [\klebanov ] it was
assumed that the fields were
bosonic but could just as well have been fermionic. The physical
states are gauge invariant and
consist of occupied closed paths of links. Each string bit carries a
single $\epsilon$ unit of
longitudinal momentum. The number of string bits in a closed string
is given by

$$
n\;=\;{p_-}/{\epsilon}
\eqn\numbit
$$

A closed configuration of string bits can be mapped to the periodic
$\sigma$ parameter space. Each
string bit corresponds to an interval $\delta{\sigma}$ of size
$\epsilon$. The range of $\sigma$
is from zero to $p_-$. A given closed string defines a discretized
version of a string
configuration $X_{\perp}(\sigma)$.

3) The dynamics is provided by a Hamiltonian containing terms both
quadratic and quartic in the
matrix fields. The quartic terms are similar to plaquette terms in
lattice gauge theory. In the
limit $N\to \infty$ the space of connected strings is closed under
the action of the hamiltonian.
An equivalent discretized string Hamiltonian describes the dynamics
and spectrum.

4) Despite the fact that the transverse lattice spacing is fixed, in
the limit $\epsilon \to 0$
and $N \to \infty$ the resulting theory exactly reproduces free
string theory. All continuous
space-time symmetries are restored and the spectrum and vertex
operators exactly reproduce
continuum free string theory.

5) For $N$ not equal to infinity there are splitting and joining
interactions similar to those in
string perturbation theory. The string coupling constant is given by

$$
g\;=\;{1\over N}
\eqn\gee
$$

It was not proven in [\klebanov ] that the $1/N$ expansion produces
string perturbation theory But
judging from similar examples, it probably does. Once again to get
exact agreement
the limit
$\epsilon \to 0$ is necessary. Unfortunately no supersymmetric
version exists so it is not at
the moment possible to eliminate the tachyon instability.

The theory described above is obviously in many ways similar to the
holographic theory described
in earlier sections. However the most important element has not been
discussed yet. What happens
to string theory when the string density gets large? It is to be
emphasized that this will
happen, even for just a single particle, as the mode cutoff
increases. Furthermore, no matter
how small we make $g$, the density eventually increases to the point
where interactions become
important.

There is no question that when the number of string bits passing
through a link becomes of order
$N^2$ interactions become very strong. This is most obvious in the
case of fermionic string bits.
The maximum number of fermionic string bits kinematically allowed on
a link is $N^2$. An effective
infinite repulsion must prevent more than this number of string bits
on a link. Just this
number of string bits per link corresponds to one bit per Planck
area.

In the bosonic case nonlinear effects do become important and it is
possible at this point that it
becomes energetically prohibitive to further increase the density. If
it can be shown that the
string density can not exceed $1/{l^2_p}$ and that the good features
of lattice string theory survive
beyond perturbation theory ( supersymmetry will surely be important
here ) then lattice string theory
is a concrete example of a holographic theory. I believe that this is
an extremely important problem
for string theorists who want to show that string theory really
exists and defines a quantum theory
of gravity.

\chapter {Information Spreading}

Let us now consider a system composed of several ordinary
particles  separated by some large transverse distance of
order$L$. As ${p_-(tot)}\over{\epsilon}$ increases to
${L^2}\over{l^2_p}$ the individual particles grow until they
begin to overlap. If we further increase the momentum, one of
two things may happen. The action of the boost operator could
conceivably lead to an increasing transverse separation of the
particles so that each maintains its spatial integrity. The
other extreme is that the spatial identity of the individual
particles is lost and they merge into a single disc whose area
is proportional to $p_-(tot)$. This I believe is the far more
reasonable answer. Before arguing in favor of this
possibility I will describe the proposal a bit more fully.

First take a single particle and boost it to extreme momentum
so that its area is macroscopic. Now consider a patch of the
screen inside the image disc. The patch may also be
macroscopic but it is assumed significantly smaller than the disc.
The region within the patch may be described by a density
matrix $\rho(patch)$. We now wish to know how much
information about the particular state of the particle is
contained in  $\rho(patch)$. As we shall see the only answer
which is consistent with what we know about black holes is
there is essentially no information in  $\rho(patch)$. In
other words the identity of the particle gets lost unless we
look at the global and very complex correlations involving
the whole disc. This kind of behavior has been brilliantly
analyzed by D. Page [\page ].

Now consider several particles which are so extremely
boosted that their image discs  completely lose their
individuality and are replaced by one large disc. Once again
I assume that the state of the system can not be recovered
by local measurements involving small patches. Two
arguments, both involving black holes, will now be given
to support these assumptions.

First, let us consider an experimental arrangement to
resolve a pair of well separated particles moving along the
$z$ axis with transverse separation $L$. A target particle at
rest is introduced. As in sect.3 the energy of the target
determines the cutoff $\epsilon$. At first if the momentum is
relatively low the target collides with at most, one of the
particles, the other continuing on its way. To see the second
particle the target must be moved.

Now let $p_-(tot)$become much larger than
${\epsilon}{L^2}\over{l^2_p}$. In this event the black hole
which forms will generally swallow both particles and the
pair will behave like a single particle. Furthermore, from
the resulting quasithermal collision products it will be very
difficult to reconstruct the original system.

One can obtain more insight into the local properties of the
image disc by considering the matter falling toward a horizon.
As we have seen, the evolution is generated by the boost
operator. As we know, the local properties of the stretched
horizon of a black hole  are described by a thermal density
matrix. Such a local thermal state has lost almost all
information about the details of the infalling matter.
Therefore it must be a property of the boost operator that it
locally obliterates the distinctions between states.

We can now imagine the following gedanken calculation to
confirm this picture. We consider a single physical particle
composed of ${p_-}\over{\epsilon}$ string bits. For
definiteness we can use the yet to be discovered lattice version of
interacting supersymmetric light front
string theory. Now let $p_- $ grow enormous and consider a
small but macroscopic patch centered near the center of mass
of the particle. First we would like to see that
${\rho}(patch)$ becomes independent of the initial particle.
Moreover the relation with the black hole horizon suggests a
quantitative test. Given the density matrix, an entropy for
the patch may be computed

$$
S\;=\;-Tr{\rho}(patch)Log[{\rho}(patch)]
$$

This should tend to the universal horizon entropy

$$
S\;=\;{{area}\over{4G}}
$$

\chapter{Discussion}

Perhaps the most disturbing feature of the theory I have
outlined is the is the continual presence of the
parameter $\epsilon$. One should of course expect that
this parameter eventually disappears from the ordinary, on
shell, $S$ matrix elements that are usually considered the
proper business of string theory. We may of course
declare that this is the only allowable business. In
so doing, with a single blow we not only eliminate
reference to $\epsilon$ but we also solve the great
paradox of black hole quantum mechanics. Simply put, in
string theory the formation and evaporation of a black
hole must be described by an $S$ matrix because string
theory is an $S$ matrix theory. Obviously this is an
unsatisfactory conclusion to the debate. In particular it
would not permit us to discuss physics from the viewpoint
of a freely falling observer, falling into a black hole.
This is a severe limitation indeed since we ourselves may
well be entering the horizon of an enormous black hole.
We have no way of knowing.

The parameter $\epsilon$ is the price we pay for a more
detailed space time description which allows any
discussion of these sorts of issues. Furthermore this
parameter is not without physical meaning. As Bohr
emphasized the discussion of a physical process is
incomplete without a specification of the apparatus. In
particular the time resolution of the apparatus plays a
central role in any effort to describe a theory which has
arbitrarily high frequency internal motions as string
theory has. The apparatus in question may be a particle
detector, an atom, or the Hawking radiation emitted from
a black hole's horizon.

Starting with the Bekenstein 't Hooft view of information and area we
have been led to an extremely
unconventional view of the quantum theory of gravity. In particular
all those concepts which
derive from the idea of a gravitational field independently
fluctuating in each Planckian volume
element have no place in the present theory. For example, the idea of
Planck scale space-time foam
or more generally, the notion of a path integral over a 3+1
dimensional metric may make no sense.
In fact it is unclear that there is any domain of phenomena in which
the conventional
gravitational field should be treated as something to be integrated
over at least with anything
like the usual integrand. At best there might be a region between the
string scale and the
Planck scale where a semiclassical approximation to a weakly
fluctuating gravitational field might
hold. It is clear that if the present ideas are correct then by the
time that the planck scale is
reached, the number of nonredundant degrees of freedom is infinitely
smaller than what is
ordinarily envisioned. It seems to me that this must have an
important long range effect on
gravitational research.

In the short term there are a number of areas in which progress can
be made. First a better
understanding of the concept of fixed points and their relation to
Lorentz boosts and high energy
scattering is possible in ordinary field theories like QCD
[\wilsonperry ].

Secondly, for  string theory to provide an interesting candidate for
a holographic theory,
several missing ingredients have to be filled in. The
first would be to show that the $1/N$ expansion reproduces the
bosonic string perturbation
expansion. However, even if this can be done, the theory can not be
analyzed nonperturbatively
because of the tachyon instability. Therefore it is very important to
determine if the lattice
model can be supersymmetrized. Assuming this is possible, we can then
attack the nonperturbative
issues. From the present point of view the most important questions
would be the following:

Does the theory retain its good features nonperturbatively. In
particular are the continuous
symmetries such as translation, rotation, and full Lorentz symmetry
left intact?

In the limit of vanishing cutoff, is the density of string bounded at
about $N^2$ bits per link?

If these questions can be answered in the affirmative then string
theory provides us with a
concrete example of a holographic world.

\ack

During the month of July of this year I visited Gerard 't Hooft in
Utrecht. Much of my
thinking about these problems was stimulated by our discussions
during this time. The idea that the
world is in a sense two dimensional is 't Hooft's but it resonated
very closely with my own
thinking about horizons and strings. The way of implementing the
idea, and its possible
connections with string theory, is mine but it was very heavily
influenced by our conversations.

I would like to thank both 't Hooft and John Preskill for emphasizing
that information can and
should spread at the causal limit as matter approaches a horizon.

Finally I benefitted from discussions with Kenneth Wilson and Robert
Perry, about
boosts and renormalization fixed points in light front quantum
mechanics and Lev Lipatov about high
energy scattering.

\refout
\end